\documentclass[letter]{aa} 

\usepackage{graphicx}
\usepackage{txfonts}
\usepackage{caption}
\usepackage{subcaption}
\usepackage{placeins}
\usepackage{cuted}
\usepackage{float}
\definecolor{cadmiumgreen}{rgb}{0.0, 0.42, 0.24}

\begin{document}

  \title{Reflections on nebulae around young stars}

  \subtitle{A systematic search for late-stage infall of material onto Class II disks}

    \author{
    A. Gupta\inst{1}\fnmsep\thanks{Aashish.Gupta@eso.org}
    \and
    A. Miotello\inst{1}
    \and
    C. F. Manara\inst{1}
    \and
    J. P. Williams\inst{10}
    \and
    S. Facchini\inst{4}
    \and
    G. Beccari\inst{1}
    \and
    T. Birnstiel\inst{2,3}
    \and
    C. Ginski\inst{5,6}
    \and
    A. Hacar\inst{7}
    \and
    M. Küffmeier\inst{8,9}
    \and
    L. Testi\inst{12}
    \and
    L. Tychoniec\inst{1}
    \and
    H.-W. Yen\inst{11}
    }
    
          
    \institute{European Southern Observatory, Karl-Schwarzschild-Str. 2, 85748 Garching bei München, Germany
    \and
    University Observatory, Faculty of Physics, Ludwig-Maximilians-Universität München, Scheinerstr. 1, 81679 Munich, Germany
    \and
    Exzellenzcluster ORIGINS, Boltzmannstr. 2, D-85748 Garching, Germany
    \and
    Universit\`a degli Studi di Milano, via Giovanni Celoria 16, 20133 Milano, Italy
    \and
    Leiden Observatory, Leiden University, Niels Bohrweg 2, NL-2333 CA Leiden, The Netherlands
    \and
    Anton Pannekoek Institute for Astronomy (API), University of Amsterdam, Science Park 904, 1098 XH Amsterdam, The Netherlands
    \and
    Department of Astrophysics, University of Vienna, Türkenschanzstrasse 17, 1180 Vienna, Austria
    \and
    Department of Astronomy, University of Virginia, Charlottesville, VA 22904, USA
    \and
    Max-Planck Institute for Extraterrestrial Physics, Gießenbachstraße 1, 85748 Garching, Germany
    \and
    Institute for Astronomy, University of Hawaii, Honolulu, HI 96822, USA
    \and
    Academia Sinica Institute of Astronomy and Astrophysics, 11F of Astro-Math Bldg, 1, Sec. 4, Roosevelt Rd, Taipei 10617, Taiwan
    \and
    Alma Mater Studiorum Università di Bologna, Dipartimento di Fisica e Astronomia (DIFA), Via Gobetti 93/2, I-40129, Bologna, Italy
    }



 
  \abstract
    {While it is generally assumed that Class II sources evolve largely in isolation from their environment, many still lie close to molecular clouds and may continue to interact with them. This may result in late accretion of material onto the disk that can significantly influence disk processes and planet formation.}  
  {In order to systematically study late infall of gas onto disks, we identify candidate Class II sources in close vicinity to a reflection nebula (RN) that may be  undergoing this process.
  }
  {First we targeted Class II sources with known kilo-au scale gas structures -- possibly due to late infall of material -- and we searched for RNe in their vicinity in optical and near-infrared images. Second, we compiled a catalogue of Class II sources associated with RNe and looked for the large-scale CO structures in archival ALMA data. 
  Using the catalogues of protostars and RNe, we also estimated the probability of Class II sources interacting with surrounding material.
  }
  {All of the sources with large-scale gas structures also exhibit some reflection nebulosity in their vicinity. Similarly, at least five Class II objects associated with a prominent RNe, and for which adequate ALMA observations are available, were found to have spirals or stream-like structures which may be due to late infall. We report the first detection of these structures around S CrA. 
  }
  {Our results suggest that a non-negligible fraction of Class II disks in nearby star-forming regions may be associated with RNe and could therefore be undergoing late accretion of gas. Surveys of RNe and kilo-au scale gas structures around Class II sources will allow us to better understand the frequency and impact of late-infall phenomena.
  }


  \maketitle
%

\section{Motivation}
\label{sec:motivation}

Most stars are born in groups in giant molecular clouds through the gravitational collapse of dense molecular cores \citep{McKee2007}. 
As these protostellar systems evolve from the Class 0/I to the Class II stage, the surrounding gaseous envelope is thought to be completely dispersed and, traditionally, Class II systems are believed to evolve in general isolation to form planetary systems.

However, in reality, these systems are still in the vicinity of molecular clouds on large scales ($\gtrsim1$~pc) and may continue to dynamically interact with them. Such interactions were observed in younger Class 0/I objects, where 1000 au scale streamers of molecular gas infalling onto the protostar were detected \citep[e.g.][]{Pineda2020,Garufi2022}. There have been some recent serendipitous detections of $\sim1000$~au, generally stream-like, gaseous structures around Class II disks too, which are likely due to such interactions \citep{Tang2012,Ginski2021,Huang2020,Huang2021,Huang2022}.

Late infall of material can greatly influence the physical and chemical properties of Class II disks, and thus, of the planets they form. For example, the supply of fresh material can help solve the `mass-budget problem' of planet-forming disks
\citep[e.g.][]{Manara2018,Mulders2021}, 
and explain the observed chemical diversity among meteorites \citep{Nanne2019N}. 
\cite{Thies2011} and \citet{Kuffmeier2021} demonstrated that late infall can torque disks and explain the observed misalignment of some planetary systems. 
Approximated as Bondi-Hoyle accretion \citep{Bondi1944}, late infall can explain the steep dependence of mass accretion on stellar mass
\citep{Padoan2005,Padoan2014}. 
Finally, this phenomenon may also produce disk sub-structures and instabilities as seen in
vortices \citep{Bae2015,Kuznetsova2022}, spiral waves \citep{Hennebelle2017,Kuffmeier2018}, and FU Orionis outbursts \citep{Dullemond2019}.

A characterisation of the frequency and efficiency of late infall is therefore crucial for establishing a holistic view of the star and planet formation process. Simulations suggest that a signpost of such accretion events could be $\sim10^3$~au-scale arc-shaped structures, generally referred to as `streamers' \citep[e.g.][]{Kuffmeier2020}. However, only a few have been detected, mainly because these spatial scales lie at the limits of single-dish resolution and they are largely filtered out in interferometric observations designed to resolve protoplanetary disks. To comprehensively study late-stage infall, a survey of large-scale structures around Class II sources is needed and a first step in this direction would be to systematically identify suitable targets.

In order to find disks that are potentially undergoing late infall, one should first identify Class II sources that are close enough to clouds to gravitationally interact with them. Such clouds
will scatter the protostellar light in optical and near-infrared (NIR) wavelengths and appear as reflection nebulae (RNe) \citep[e.g.][]{Hubble1922}. Historically, RNe were used to identify young stellar objects (YSOs)
\citep[e.g.][]{Cohen1980} and
they were   indeed one of the original defining characteristics of T Tauri stars \citep{Joy1945}.
Recently, hydrodynamical simulations have demonstrated that kilo-au scale RNe can appear due to cloud-protostar interactions, some of which lead to late infall of material \citep{Dullemond2019}. In this Letter we pioneer the use of RN detections close to Class II stars to identify late-infall candidates. 

\section{Class II sources with large-scale CO structures}
\label{sec:knownsources}

\begin{figure}
\centering
\includegraphics[scale=0.70]{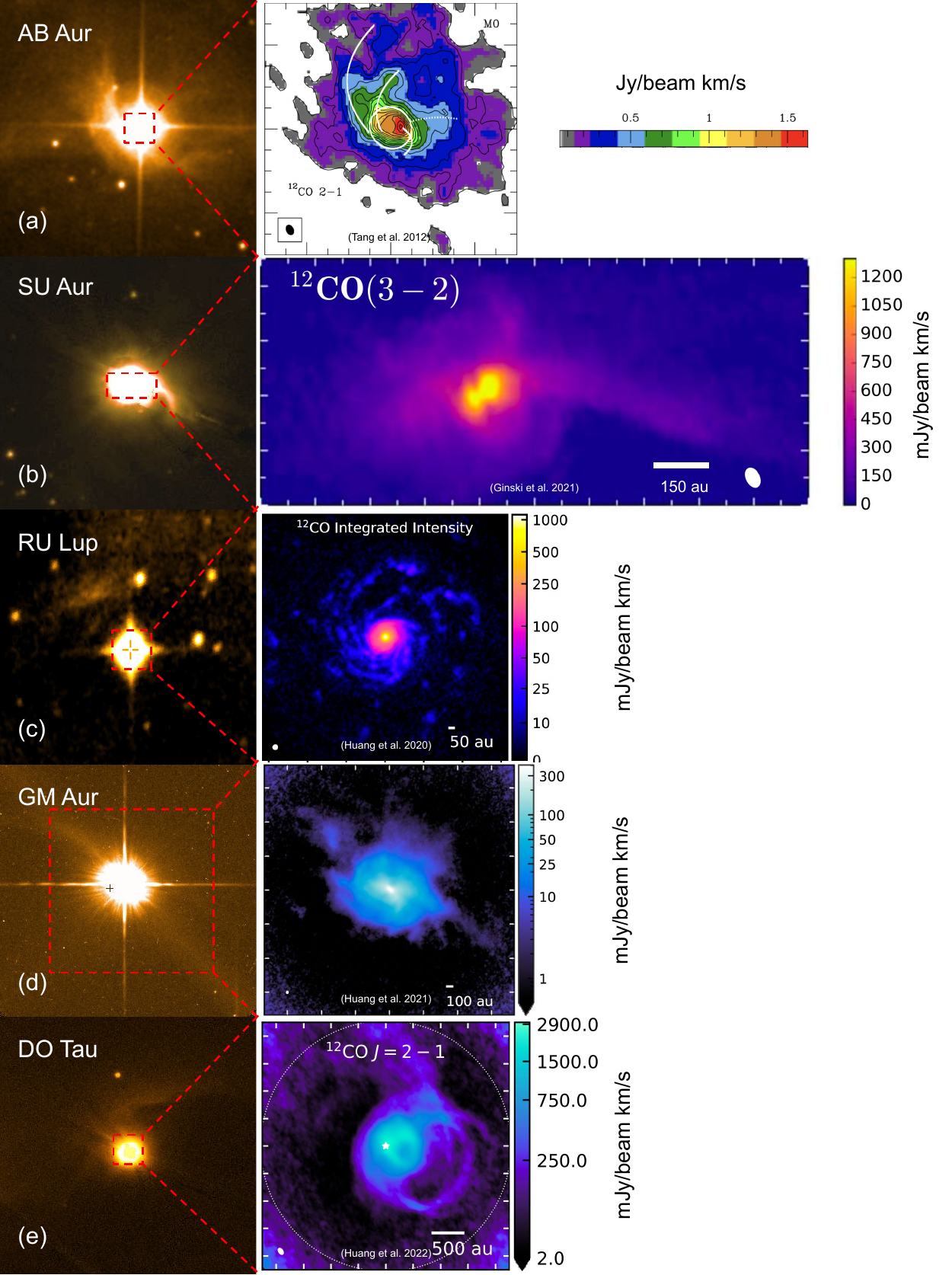}
\caption{Optical and NIR images (left) and $^{12}$CO integrated intensity moment 0 maps (right) of Class II sources with known large-scale structures, as discussed in Section \ref{sec:knownsources}. 
{\it Panel a}: AB Aur's DSS2 optical image and the PdBI $^{12}$CO (2--1) moment 0 map from \citet{Tang2012}. 
{\it Panel b}: SU Aur's Pan-STARRS optical image and the ALMA $^{12}$CO (3--2) moment 0 map from \citet{Ginski2021}.
{\it Panel c}: RU Lup's DSS2 (red) optical image and the ALMA $^{12}$CO (2--1) moment 0 map from \citet{Huang2020}.
{\it Panel d}: GM Aur's HST (NICMOS) NIR image and the ALMA $^{12}$CO (2--1) moment 0 map from \citet{Huang2021}.
{\it Panel e}: DO Tau's Pan-STARRS optical image and ALMA $^{12}$CO (2--1) moment 0 map from \citet{Huang2022}.
Angular resolutions for the moment 0 maps are roughly 0.5\arcsec, 0.3\arcsec, 0.3\arcsec, 0.2\arcsec, and 0.7\arcsec\  for panels a to e, respectively.
}
\label{fig:known_sources}
\end{figure}

Large-scale non-Keplerian gaseous structures have been detected around a few Class II sources. In order to test the hypothesis that RNe might indicate late infall,
we looked for signs of nebulosity around these sources, as discussed below:

\smallskip
\emph{AB Aur:} \citet{Tang2012} found four $\sim500$~au spirals around AB Aur in CO observations using the Instituto de Radioastronomía Milimétrica (IRAM) 30-m telescope, Plateau de Bure interferometer (PbDI), and Submillimeter Array (SMA). Analysis of the gas kinematics in these spirals, which seems to be counter-rotating with respect to the Keplerian disk, suggests that they are likely formed due to late-stage infall of gas. This source is also known to be associated with a bright arc-shaped RN \citep[e.g.][]{Dullemond2019}, as shown in Figure \ref{fig:known_sources} (panel a).
    
\smallskip
\emph{SU Aur:} \citet{Akiyama2019} reported an $\sim1000$~au long tail-like streamer using the Atacama Large Millimeter/submillimeter Array (ALMA) in CO emission. Later, \citet{Ginski2021} studied the morphology of its dust tails in scattered light using
the Very Large Telescope (VLT), along with a kinematic study of CO gas, and found that the material is likely moving towards the disk. A bright RN is visible in the immediate vicinity of SU Aur (Figure \ref{fig:known_sources}, panel b).
    
\smallskip
\emph{RU Lup:} \citet{Huang2020} reported at least five CO spiral arms stretching up to $\sim1000$~au around RU Lup using ALMA observations and suggested late infall as a possible explanation. Archival Digital Sky Survey (DSS) images show faint nebulosity just north of RU Lup, as shown in Figure \ref{fig:known_sources} (panel c).

\smallskip
\emph{GM Aur:} \citet{Huang2021} found extended elongated structures around GM Aur, $\sim1000$--2000~au in length, using ALMA CO observations with a morphology and kinematics indicative of late infall.
No RNe have been found in archival DSS or Panoramic Survey Telescope and Rapid Response System (Pan-STARRS) images, but there are elongated features in more sensitive Hubble Space Telescope (HST) NIR images, as reported in \citet{Schneider2003} and shown in Figure \ref{fig:known_sources} (panel d).
    
\smallskip
\emph{DO Tau:} \citet{Huang2022} studied the kilo-au environment of DO Tau using VLT/SPHERE, HST, and ALMA observations, and found the disk to be connected to multiple $\sim1000$~au~scale stream-like structures. Larger-scale Herschel observations show that these structures are probably due to an interaction with the neighbouring YSO HV Tau, but late accretion of material onto DO Tau is not ruled out. The source is associated with a prominent RN (Figure \ref{fig:known_sources}, panel e).

\smallskip
To summarise, the five known Class II sources that have large-scale CO structures suggestive of an interaction with surrounding gas have either a prominent RN or some hints of reflection nebulosity. 
Following this lead,
we exploit the association of YSOs with RNe to search for candidate Class II sources undergoing late-stage infall of material.


\section{Class II sources near reflection nebulae}
\label{sec:RNesources}

As a first step, we compiled a catalogue of RNe using the published lists in \citet{Magakian2003} and \citet{Connelley2007}. \citet{Magakian2003} merged previously published RNe catalogues and presented a final list of 913 objects. Most of the sources in their list have been manually identified by visual inspection of DSS optical images. \citet{Connelley2007} surveyed 197 nearby, mostly Class I, protostars at $2.2\,\mu$m using the University of Hawaii 2.2\,m telescope. They detected 106 RNe, out of which 41 were reported as new discoveries.
The fact that $\gtrsim 40\%$ of RNe were not detected before suggests that the previous RNe catalogues were rather incomplete.
For example, the prominent RNe around SU Aur (see Figure \ref{fig:known_sources}, panel b) and HD 100546 \citep{Ardila2007} were not in these RNe catalogues.

\citet{Connelley2007} estimated sizes of RNe as the square root of their area
with a mean ($\mu$) and standard deviation ($\sigma$) of $\sim18\arcsec$ and $\sim15\arcsec$, respectively. Using a cross-matching radius of $30\arcsec$ ($\sim\mu+1\sigma$), we found ten RNe present in both of the catalogues. After removing duplicates,
we merged the two catalogues to have a final sample of 1009 RNe.

To avoid spurious detections and have a well-characterised sample of YSOs with RNe that can be further followed up on using molecular-line observations (as discussed in Appendix \ref{sec:discussion_future}), we focus
on nearby well-studied star-forming regions (SFRs). The regions considered in this study are listed in Table \ref{tab:sfrs} and shown in Figure \ref{fig:DSSmaps}. The radii reported in the fifth column in Table \ref{tab:sfrs} were used to define circular boundaries around the central coordinates listed in the second and third columns, and they are marked with black circles in Figure \ref{fig:DSSmaps}.
We found 141 sources from our catalogue of 1009 RNe within these region boundaries.

The next step was to compile a catalogue of YSOs in these regions. For this we started with the all-sky catalogue of 133980 Class I/II sources reported in \citet{Marton2016}. \citet{Marton2016} analysed \textit{WISE} and \textit{2MASS} photometry of these sources, using the support vector machine algorithm to identify them as YSOs. For a more complete sample of nearby YSOs, we also used the list of 2966 YSOs identified using \textit{Spitzer}'s 'cores to disks' and 'Gould Belt' surveys, as reported in \citet{Dunham2015}. Among the SFRs considered in this study, Taurus and Orion were not part of the \citet{Dunham2015} catalogue. Using a cross-matching radius of $5\arcsec$, we found 781 common sources in both datasets. Our final catalogue consists of 136165 YSOs of which 4930 lie within the SFR boundaries.

\begin{table}
\tiny
\caption{List of SFRs considered in this study.}
\begin{tabular}{ccccccc}
\hline\hline
SFR & R.A. & Decl. & Distance & Radius & YSOs & RNe \\
 & [deg] & [deg] & [pc] & [deg] &  &  \\
\hline
Ophiuchus & 249.07 & -23.64 & 128.0 & 6.0 & 450 & 13 \\
Taurus & 67.11 & 26.62 & 148.0 & 8.0 & 190 & 28 \\
Corona Australis & 287.96 & -38.11 & 155.0 & 4.0 & 80 & 6 \\
Lupus & 240.08 & -36.56 & 158.0 & 8.0 & 480 & 10 \\
Chamaeleon & 169.5 & -78.17 & 190.0 & 8.0 & 148 & 7 \\
Perseus & 54.21 & 31.57 & 284.0 & 4.0 & 435 & 19 \\
Orion & 85.2 & -3.5 & 420.0 & 6.0 & 978 & 44 \\
Serpens & 277.66 & -1.91 & 495.0 & 3.0 & 2169 & 14 \\
\hline
\end{tabular}
\tablefoot{The median coordinates and distances are from \citet{Zucker2020}. The last three columns refer to the radii used to define the SFR boundaries (see Figure \ref{fig:DSSmaps}) and the number of RNe and YSOs found.}
\label{tab:sfrs}
\end{table}

The spectral indices, $\alpha$, or slope of the spectral energy distributions in the infrared regime ($\sim2$--$25~\mu$m) were given in the \citet{Dunham2015} catalogue.
We used their values, $\alpha'$, corrected for foreground extinction to classify the source evolutionary state.

Similarly, we determined $\alpha$ for sources exclusively in the \citet{Marton2016} catalogue using the provided \textit{WISE} and \textit{2MASS} (K band) photometry, and discarding those with fitting uncertainties $>0.6$, that is roughly half of the range of values for Class II sources. 
In order to estimate extinction-corrected spectral indices ($\alpha'$), we computed a correction factor of -0.31 as the median of differences between all the $\alpha'$ and $\alpha$ values reported in \citet{Dunham2015}. 
We note that this method only provides a rough estimate of $\alpha'$ as it does not account for extinction values for individual sources. 
A comparison of our $\alpha$ and $\alpha'$ values is shown in Figure \ref{fig:alphas}. Following the same classification criteria as \citet{Dunham2015}, that is $-1.6 \leq \alpha' < -0.3$, 2562 out of 4930 YSOs in our SFRs were classified as Class II sources. 

The next step was to cross match our list of Class II sources with the merged catalogue of RNe.
For the specific goal of this work,
the cross-matching distance between a YSO and RN
can be empirically estimated as the length scale from where ambient gas can be accreted onto an isolated star \citep[Bondi-Hoyle accretion,][]{Bondi1944,Throop2008,Padoan2014}. The typical length scale for this interaction is given as $L_{BH} = 2GM_{*}/v^{2}$, where $G$ is the gravitational constant, $M_{*}$ is the stellar mass, and $v$ is the stellar speed relative to the gas. For a stellar mass of 1~M$_{\odot}$ and typical stellar velocity of 1~km~s$^{-1}$, $L_{BH}\sim2000$~au. 
The 21 late-infall Class II candidates that we found using this threshold are listed in Appendix \ref{app:table}.

We note that this distance threshold is a lower limit for an interaction between clouds and YSOs because we do not account for the sizes of RNe.
These show a range of physical sizes ($\sim10^3$--$10^5$~au) and have highly asymmetrical shapes \citep[e.g.][]{Connelley2007}. When mining the ALMA archive, we ignored this diversity as we aimed to identify a reliable sample of YSOs potentially interacting with RNe clouds. The implications of a more realistic distance threshold is discussed in Section \ref{sec:discussion_RNe}.

\begin{figure}[htp]
\includegraphics[scale=0.62]{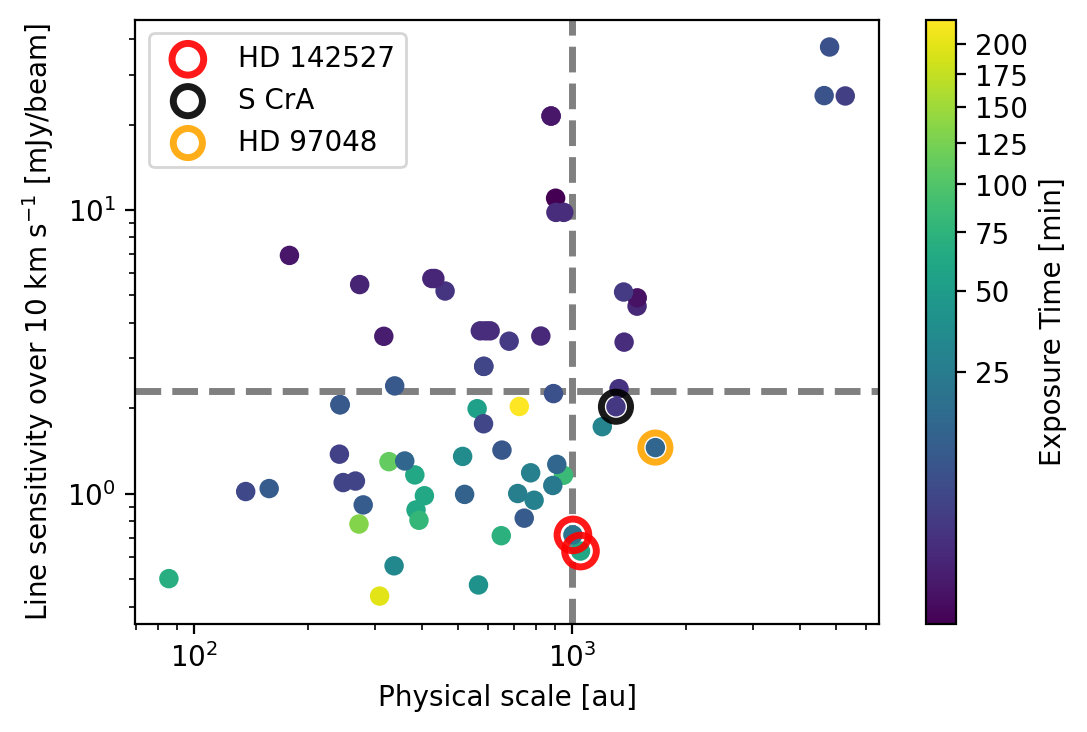}
\caption{Largest recoverable physical scale vs line sensitivity (over $10$~km~s$^{-1}$) for archival ALMA observations, in Band 6 and 7, of the 16 nearby Class II YSOs associated with RNe, as discussed in Section \ref{sec:RNesources}. Marker colours denote the exposure time for each observation in minutes. The vertical dashed line denotes a recoverable scale of 1000 au. The horizontal dashed line denotes a line sensitivity of $2.3$~mJy~beam$^{-1}$ (the line sensitivity reported in the ALMA archive for SU Aur observations by \cite{Ginski2021}). Markers with open circles denote the observations discussed in Section \ref{sec:RNesources}, with red circles for HD 142527, black circle for S CrA, and an orange circle for HD 97048.}
\label{fig:almaobs}
\end{figure}

\paragraph{ALMA archive search:} To target higher-quality observations, we focussed on nearby SFRs ($d \lesssim$ 200 pc), where 16 out of the 21 Class II disks near RNe are located.
For these nearby sources, we searched for existing observations in the ALMA Science Archive. 
We are primarily interested in $^{12}$CO emission ($J=2-1$ in Band 6 and $J=3-2$ in Band 7)
since this molecule is expected to be a good tracer of large-scale diffuse gas structures (see e.g. Figure \ref{fig:known_sources}).  
We found a total of 66 Band 6 and 7 observations for 16 sources in our sample, as shown in Figure \ref{fig:almaobs}. 

Among these datasets, we found 13 observations with the largest angular scale (LAS) corresponding to at least 1000 au to recover the expected large-scale structures \citep[see][]{Kuffmeier2020}.
As such large-scale structures are expected to have low column densities, and therefore faint emission, the observational sensitivity is important. We selected a subset of five observations with rms noise (over a channel width of $10$~km~s$^{-1}$) smaller than a $2.3$~mJy~beam$^{-1}$, equal to the line sensitivity reported in the ALMA archive for SU Aur observations \citep{Ginski2021}. 
For the typical angular resolution of $0.7\arcsec$ for these observations, the sensitivity threshold of the $2.3$~mJy~beam$^{-1}$ corresponds to $\sim100$~mK.
The sensitivity and recoverable scale thresholds are marked as dashed-grey lines in Figure \ref{fig:almaobs}.

\begin{figure*}
\centering
\includegraphics[scale=0.950]{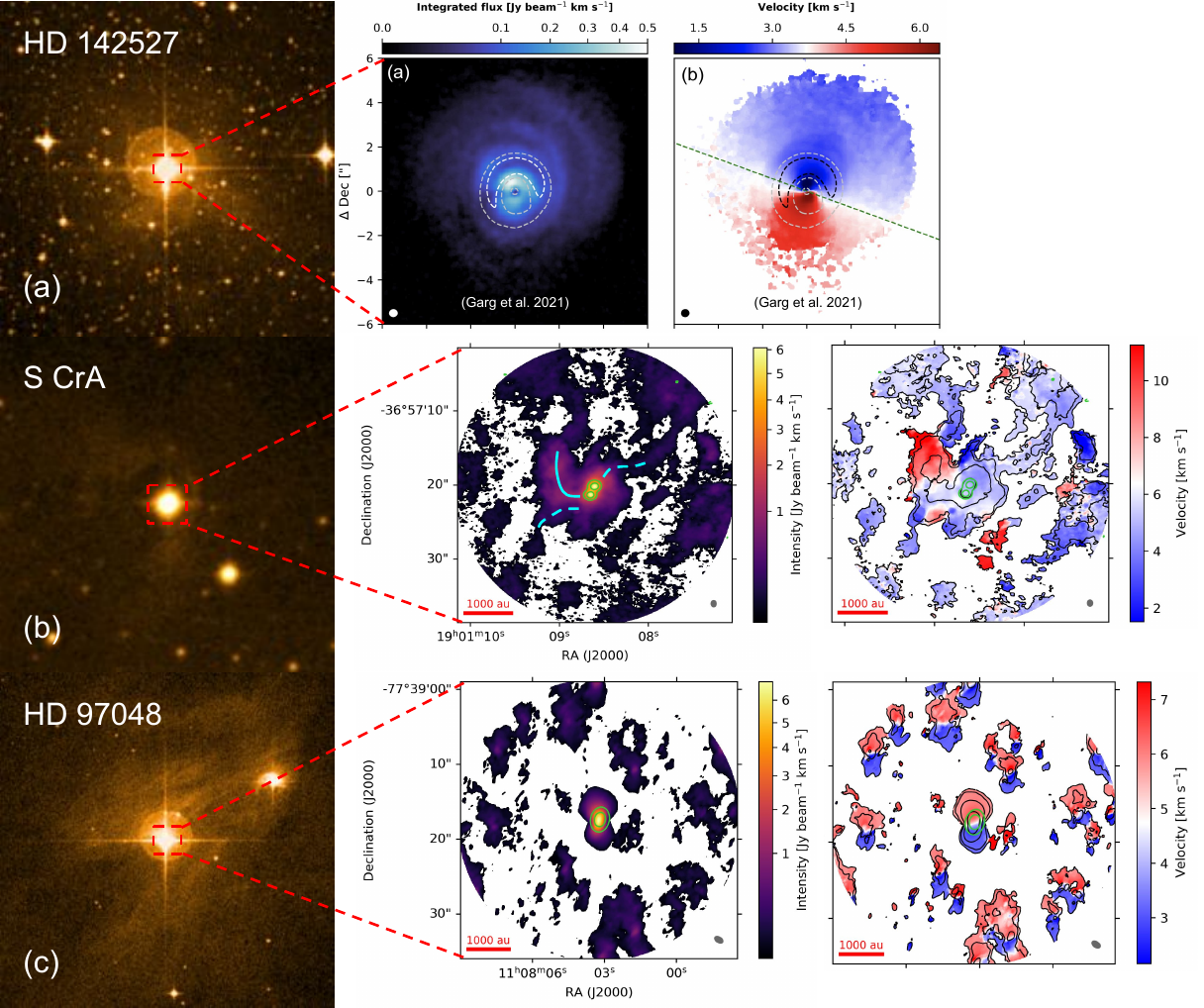}
\caption{Optical images (left), $^{12}$CO integrated intensity (moment 0) maps (middle), and $^{12}$CO intensity-weighted velocity (moment 1) maps (right) of nearby Class II sources associated with RNe, as discussed in Section \ref{sec:RNesources}. 
{\it Panel a}: HD 142527's DSS optical image and the ALMA $^{12}$CO (2--1) moment maps from \citet{Garg2021}. 
{\it Panel b}: S CrA's DSS2 (red) optical image and the ALMA $^{12}$CO (2--1) moment maps. Solid and dashed curved-cyan lines denote prominent and potential streamer-like features, respectively.
{\it Panel c}: HD 97048's DSS optical image and the ALMA $^{12}$CO (2--1) moment maps.
For the moment maps (middle and right) in panel b and c, only pixels with an intensity $>3\sigma$ are considered. In these panels, green contours represent continuum emission ($\sim1.3$~mm, 3$\sigma$ and 15$\sigma$ levels) from protoplanetary disks, horizontal red lines in the bottom-left corner represent a 1000 au length scale, the grey ellipse in the bottom-right corner represent the beam size, and black contours in moment 1 maps (right) represent moment 0 emission (starting from the error in moment 0, increased by a factor of five). Errors in moment 0 emission are 22.1 and 20.6 mJy~beam$^{-1}$~km~s$^{-1}$ for S CrA and HD 97048, respectively.}
\label{fig:rne_sources}
\end{figure*}

Among these five observations, two targeted HD 142527 (Figure \ref{fig:almaobs}, red circles), a well-studied binary Class II system.
These data have been published and show
non-Keplerian spiral structures extending to $\sim700$~au, beyond the disk radius of $\sim200$--$300$~au \citep{Christiaens2014,Garg2021}, as shown in Figure \ref{fig:rne_sources} (panel a). Similar structures have also been observed in optical and NIR images \citep{Casassus2012,Hunziker2021}. \citet{Christiaens2014} suggested that the innermost spiral can be explained by acoustic waves due to an embedded companion; however, the origin of the outer spirals is less clear, and stellar encounters and gravitational instabilities have been suggested as possible causes. We note that spiral structures have been predicted to form due to late infall \citep{Hennebelle2017,Kuffmeier2017,Kuffmeier2018} and a detailed kinematic analysis can be done to distinguish among these scenarios, as discussed in Appendix \ref{sec:discussion_future}. HD 142527 also exhibits inner and outer disk misalignment \citep{Bohn2022}, which can be explained by late infall \citep{Thies2011,Kuffmeier2021}.

The other three datasets are of S CrA, HD 97048, and Sz 68. The selected large-scale observation for Sz 68 (Project 2019.1.01135.S) does not cover any CO lines and is therefore not discussed further. For S CrA (Project 2019.1.01792.S) and HD 97048 (Project 2015.1.00192.S), we used the standard pipeline calibration and imaged the $^{12}$CO (2--1) data using CASA 6.4. For both datasets, imaging was carried out using 'briggs' weighting with robust=0.5, a cell size of 0.05\arcsec, and 'auto-multithresh' masking with default parameters. The resulting integrated intensity (moment 0) maps and intensity-weighted velocity (moment 1) maps are shown in Figure \ref{fig:rne_sources} (panel b and c). Appendix \ref{app:channel_maps} shows corresponding channel maps for both of the datasets. 

For S CrA, at least one $\sim1000$~au streamer is clearly visible (solid cyan line), north-west of the binary disks (green contours), as seen in the moment 0 map (Figure \ref{fig:rne_sources}, panel b, middle panel). This streamer seems to be redshifted with respect to the disk emission, as seen in the corresponding moment 1 map (see Figure \ref{fig:rne_sources}, panel b, right panel and channels from 12.93 to 8.57~km~s$^{-1}$ in Figure \ref{fig:scra_channels}). Furthermore, there are hints of two more streamers, denoted as dashed-cyan lines in the moment 0 map (Figure \ref{fig:rne_sources}, panel b, middle panel). This is the first discovery of large-scale streamers around S CrA. As discussed in Section \ref{sec:knownsources}, such elongated structures can form due to the late infall of material, but they could also be a consequence of a binary interaction. Further analysis is required to ascertain the dynamical nature of these features and will follow in a future publication. 

For HD 97048, no clear streamers are visible in the moment 0 map (Figure \ref{fig:rne_sources}, panel c, middle panel). 
However, significant negative emission was observed in the central channels (Figure \ref{fig:hd97048_channels}, channels 5.32 to 3.57~km~s$^{-1}$), suggesting that the source may be surrounded by large-scale gas that may absorb or otherwise obscure, through spatial filtering, kilo-au features around the star.

These initial results suggest that at least two (HD 142527 and S CrA) out of the three Class II sources associated with RNe, and with good-enough ALMA data, exhibit large-scale spirals or streamer-like structures, as are expected to form due to cloud-disk interactions, particularly late-stage infall of material \citep{Hennebelle2017,Dullemond2019,Kuffmeier2017,Kuffmeier2020, Kuznetsova2020}. 
These two sources are in addition to the already known similar sources discussed in Section \ref{sec:knownsources}.
Other possible explanations for these large-scale structures are discussed in Appendix \ref{sec:discussion_other}.
Irrespective of the actual origin of such features, it is promising to see that large-scale gas emission is found around YSOs close to RNe. This may indicate that an association with RNe can be used to look for similar structures around a wider population of Class II disks, as discussed further in Section \ref{sec:discussion_RNe}.

\section{Discussion}
\label{sec:discussion_RNe}



We have found 21 Class II sources associated with RNe in the SFRs listed in Table \ref{tab:sfrs}, using a distance threshold of $\sim2000$~au, equivalent to the typical Bondi-Hoyle accretion length scale. However, this distance threshold does not account for the observed range of physical extents and asymmetric shapes of RNe.
\citet{Connelley2007} provided angular sizes and distances for RNe in their catalogue, which correspond to a mean radius $\sim10^4$~au.
The exact distances between the centres of YSOs and RNe that result in late infall can vary greatly for different sources depending on their stellar (mass and velocity) and RNe properties (size and shape) and it is likely that more Class II sources in our sample may be interacting with neighbouring material than discussed here.



Figure \ref{fig:OffsetDist} plots the fraction of Class II sources
and all YSOs
associated with RNe as a function of the offset distance for different SFRs. 
Considerable differences can be seen in the association probability of YSOs with RNe for different SFRs, mostly due to different catalogue completeness levels.
The incompleteness of the available RNe catalogues is a major obstacle in providing reliable statistics at the moment (see discussion in Sec. \ref{sec:RNesources}). However, even with this limitation, it is clear that there are potentially many Class II disks interacting with their parent cloud. For at least four SFRs (Taurus, Lupus, Corona Aurstralis, and Chamaeleon), $\sim5$--$10\%$ of Class II sources are close-enough ($\lesssim 10^4$~au) to RNe. If the threshold is slightly increased to $\lesssim 4\times 10^4$~au, $\sim50\%$ of Class II sources in Corona Aurstralis would be associated with a RN, and therefore potentially accreting material from the ambient cloud.
Accounting for stellar kinematics, the number of Class II systems that pass by RNe at some point in their lifetime could be even greater.
If an association with RNe is indeed related to late infall,
this may be an important phenomenon
especially since it can have important implications for disk evolution and planet formation, as described in Section \ref{sec:motivation}. To further test the tentative link between RNe and an interaction between disks and surrounding clouds, a survey of structures and kinematics around Class II sources with known RNe is needed,
as discussed in Appendix \ref{sec:discussion_future}. Coupled with a better RNe catalogue, such a survey will allow us to understand how frequent late infall is for Class II sources.

\section{Conclusions}
\label{sec:conclusions}
In this Letter we pioneer the use of the detections of RNe close to Class II stars
to identify late-infall candidates. We find that all of the sources with known large-scale CO structures, where late infall is invoked as a possible explanation, also exhibit some reflection nebulosity at OIR wavelengths.
Furthermore, at least five out of the six sources which are associated with a prominent RNe and for which adequate ALMA observations are available -- that is, known sources AB Aur, SU Aur, and DO Tau along with independently identified sources S CrA and HD 142527 -- exhibit some large scale structure that may be indicative of late infall.
This per se suggests that association with RNe may be used to identify candidate Class II sources undergoing late-stage infall of material. Finally, in nearby SFRs, the fraction of Class II sources associated with RNe can be as large as $50\%$, depending on the distance threshold, but a proper statistical analysis is still pending improved RNe catalogues. If RNe are indeed related to late infall, this suggests that a significant fraction of Class II sources could be undergoing this phenomenon, with a non-negligible impact on disk evolution and planet formation.
The catalogue of potential late accretors obtained serves as a starting point for more systematic studies of late infall onto disks.

    
        
    

\begin{acknowledgements}
This work was partly funded by the Deutsche Forschungsgemeinschaft (DFG, German Research Foundation) - 325594231.
 
 
Funded by the European Union under the European Union’s Horizon Europe Research \& Innovation Programme 101039452 (WANDA). 

T.B. acknowledges funding from the European Research Council (ERC) under the European Union's Horizon 2020 research and innovation programme under grant agreement No 714769 and funding by the Deutsche Forschungsgemeinschaft (DFG, German Research Foundation) under grants 361140270, 325594231 (FOR 2634/2), and Germany's Excellence Strategy - EXC-2094 - 390783311.

M.K. gratefully acknowledges that this project has received funding from the European Union’s Framework Programme for Research and Innovation Horizon 2020 (2014–2020) under the Marie Skłodowska-Curie Grant Agreement No. 897524.

This work was partly supported by the Italian Ministero dell’Istruzione, Universit\`{a} e Ricerca through the grant Progetti Premiali 2012-iALMA (CUP C52I13000140001), by the DFG Cluster of Excellence Origins (www.origins-cluster.de). This project has received funding from the European Union’s Horizon 2020 research and innovation program under the Marie Sklodowska-Curie grant agreement No 823823 (DUSTBUSTERS) and from the European Research Council (ERC) via the ERC Synergy Grant ECOGAL (grant 855130).

Views and opinions expressed are however those of the author(s) only and do not necessarily reflect those of the European Union or the European Research Council. Neither the European Union nor the granting authority can be held responsible for them.

This paper makes use of the following ALMA data: ADS/JAO.ALMA\#2015.1.00192.S and ADS/JAO.ALMA\#2019.1.01792.S. ALMA is a partnership of ESO (representing its member states), NSF (USA) and NINS (Japan), together with NRC (Canada), MOST and ASIAA (Taiwan), and KASI (Republic of Korea), in cooperation with the Republic of Chile. The Joint ALMA Observatory is operated by ESO, AUI/NRAO and NAOJ.

This work made use of Astropy:\footnote{http://www.astropy.org} a community-developed core Python package and an ecosystem of tools and resources for astronomy \citep{astropy:2013, astropy:2018, astropy:2022}.

\end{acknowledgements}

\bibliographystyle{aa} 
\bibliography{refs} 

\begin{thebibliography}{59}
\expandafter\ifx\csname natexlab\endcsname\relax\def\natexlab#1{#1}\fi

\bibitem[{{Akiyama} {et~al.}(2019){Akiyama}, {Vorobyov}, {Liu}, {Dong}, {de
  Leon}, {Liu}, \& {Tamura}}]{Akiyama2019}
{Akiyama}, E., {Vorobyov}, E.~I., {Liu}, H.~B., {et~al.} 2019, \aj, 157, 165

\bibitem[{{Ardila} {et~al.}(2007){Ardila}, {Golimowski}, {Krist}, {Clampin},
  {Ford}, \& {Illingworth}}]{Ardila2007}
{Ardila}, D.~R., {Golimowski}, D.~A., {Krist}, J.~E., {et~al.} 2007, \apj, 665,
  512

\bibitem[{{Astropy Collaboration} {et~al.}(2022){Astropy Collaboration},
  {Price-Whelan}, {Lim}, {Earl}, {Starkman}, {Bradley}, {Shupe}, {Patil},
  {Corrales}, {Brasseur}, {N{"o}the}, {Donath}, {Tollerud}, {Morris},
  {Ginsburg}, {Vaher}, {Weaver}, {Tocknell}, {Jamieson}, {van Kerkwijk},
  {Robitaille}, {Merry}, {Bachetti}, {G{"u}nther}, {Aldcroft},
  {Alvarado-Montes}, {Archibald}, {B{'o}di}, {Bapat}, {Barentsen}, {Baz{'a}n},
  {Biswas}, {Boquien}, {Burke}, {Cara}, {Cara}, {Conroy}, {Conseil}, {Craig},
  {Cross}, {Cruz}, {D'Eugenio}, {Dencheva}, {Devillepoix}, {Dietrich},
  {Eigenbrot}, {Erben}, {Ferreira}, {Foreman-Mackey}, {Fox}, {Freij}, {Garg},
  {Geda}, {Glattly}, {Gondhalekar}, {Gordon}, {Grant}, {Greenfield}, {Groener},
  {Guest}, {Gurovich}, {Handberg}, {Hart}, {Hatfield-Dodds}, {Homeier},
  {Hosseinzadeh}, {Jenness}, {Jones}, {Joseph}, {Kalmbach}, {Karamehmetoglu},
  {Ka{l}uszy{'n}ski}, {Kelley}, {Kern}, {Kerzendorf}, {Koch}, {Kulumani},
  {Lee}, {Ly}, {Ma}, {MacBride}, {Maljaars}, {Muna}, {Murphy}, {Norman},
  {O'Steen}, {Oman}, {Pacifici}, {Pascual}, {Pascual-Granado}, {Patil},
  {Perren}, {Pickering}, {Rastogi}, {Roulston}, {Ryan}, {Rykoff}, {Sabater},
  {Sakurikar}, {Salgado}, {Sanghi}, {Saunders}, {Savchenko}, {Schwardt},
  {Seifert-Eckert}, {Shih}, {Jain}, {Shukla}, {Sick}, {Simpson},
  {Singanamalla}, {Singer}, {Singhal}, {Sinha}, {Sip{H{o}}cz}, {Spitler},
  {Stansby}, {Streicher}, {{{S}}umak}, {Swinbank}, {Taranu}, {Tewary},
  {Tremblay}, {Val-Borro}, {Van Kooten}, {Vasovi{'c}}, {Verma}, {de Miranda
  Cardoso}, {Williams}, {Wilson}, {Winkel}, {Wood-Vasey}, {Xue}, {Yoachim},
  {Zhang}, {Zonca}, \& {Astropy Project Contributors}}]{astropy:2022}
{Astropy Collaboration}, {Price-Whelan}, A.~M., {Lim}, P.~L., {et~al.} 2022,
  apj, 935, 167

\bibitem[{{Astropy Collaboration} {et~al.}(2018){Astropy Collaboration},
  {Price-Whelan}, {Sip{\H{o}}cz}, {G{\"u}nther}, {Lim}, {Crawford}, {Conseil},
  {Shupe}, {Craig}, {Dencheva}, {Ginsburg}, {Vand erPlas}, {Bradley},
  {P{\'e}rez-Su{\'a}rez}, {de Val-Borro}, {Aldcroft}, {Cruz}, {Robitaille},
  {Tollerud}, {Ardelean}, {Babej}, {Bach}, {Bachetti}, {Bakanov}, {Bamford},
  {Barentsen}, {Barmby}, {Baumbach}, {Berry}, {Biscani}, {Boquien}, {Bostroem},
  {Bouma}, {Brammer}, {Bray}, {Breytenbach}, {Buddelmeijer}, {Burke},
  {Calderone}, {Cano Rodr{\'\i}guez}, {Cara}, {Cardoso}, {Cheedella}, {Copin},
  {Corrales}, {Crichton}, {D'Avella}, {Deil}, {Depagne}, {Dietrich}, {Donath},
  {Droettboom}, {Earl}, {Erben}, {Fabbro}, {Ferreira}, {Finethy}, {Fox},
  {Garrison}, {Gibbons}, {Goldstein}, {Gommers}, {Greco}, {Greenfield},
  {Groener}, {Grollier}, {Hagen}, {Hirst}, {Homeier}, {Horton}, {Hosseinzadeh},
  {Hu}, {Hunkeler}, {Ivezi{\'c}}, {Jain}, {Jenness}, {Kanarek}, {Kendrew},
  {Kern}, {Kerzendorf}, {Khvalko}, {King}, {Kirkby}, {Kulkarni}, {Kumar},
  {Lee}, {Lenz}, {Littlefair}, {Ma}, {Macleod}, {Mastropietro}, {McCully},
  {Montagnac}, {Morris}, {Mueller}, {Mumford}, {Muna}, {Murphy}, {Nelson},
  {Nguyen}, {Ninan}, {N{\"o}the}, {Ogaz}, {Oh}, {Parejko}, {Parley}, {Pascual},
  {Patil}, {Patil}, {Plunkett}, {Prochaska}, {Rastogi}, {Reddy Janga},
  {Sabater}, {Sakurikar}, {Seifert}, {Sherbert}, {Sherwood-Taylor}, {Shih},
  {Sick}, {Silbiger}, {Singanamalla}, {Singer}, {Sladen}, {Sooley},
  {Sornarajah}, {Streicher}, {Teuben}, {Thomas}, {Tremblay}, {Turner},
  {Terr{\'o}n}, {van Kerkwijk}, {de la Vega}, {Watkins}, {Weaver}, {Whitmore},
  {Woillez}, {Zabalza}, \& {Astropy Contributors}}]{astropy:2018}
{Astropy Collaboration}, {Price-Whelan}, A.~M., {Sip{\H{o}}cz}, B.~M., {et~al.}
  2018, \aj, 156, 123

\bibitem[{{Astropy Collaboration} {et~al.}(2013){Astropy Collaboration},
  {Robitaille}, {Tollerud}, {Greenfield}, {Droettboom}, {Bray}, {Aldcroft},
  {Davis}, {Ginsburg}, {Price-Whelan}, {Kerzendorf}, {Conley}, {Crighton},
  {Barbary}, {Muna}, {Ferguson}, {Grollier}, {Parikh}, {Nair}, {Unther},
  {Deil}, {Woillez}, {Conseil}, {Kramer}, {Turner}, {Singer}, {Fox}, {Weaver},
  {Zabalza}, {Edwards}, {Azalee Bostroem}, {Burke}, {Casey}, {Crawford},
  {Dencheva}, {Ely}, {Jenness}, {Labrie}, {Lim}, {Pierfederici}, {Pontzen},
  {Ptak}, {Refsdal}, {Servillat}, \& {Streicher}}]{astropy:2013}
{Astropy Collaboration}, {Robitaille}, T.~P., {Tollerud}, E.~J., {et~al.} 2013,
  \aap, 558, A33

\bibitem[{{Bae} {et~al.}(2015){Bae}, {Hartmann}, \& {Zhu}}]{Bae2015}
{Bae}, J., {Hartmann}, L., \& {Zhu}, Z. 2015, \apj, 805, 15

\bibitem[{{Bohn} {et~al.}(2022){Bohn}, {Benisty}, {Perraut}, {van der Marel},
  {W{\"o}lfer}, {van Dishoeck}, {Facchini}, {Manara}, {Teague}, {Francis},
  {Berger}, {Garcia-Lopez}, {Ginski}, {Henning}, {Kenworthy}, {Kraus},
  {M{\'e}nard}, {M{\'e}rand}, \& {P{\'e}rez}}]{Bohn2022}
{Bohn}, A.~J., {Benisty}, M., {Perraut}, K., {et~al.} 2022, \aap, 658, A183

\bibitem[{{Bondi} \& {Hoyle}(1944)}]{Bondi1944}
{Bondi}, H. \& {Hoyle}, F. 1944, \mnras, 104, 273

\bibitem[{{Casassus} {et~al.}(2012){Casassus}, {Perez M.}, {Jord{\'a}n},
  {M{\'e}nard}, {Cuadra}, {Schreiber}, {Hales}, \& {Ercolano}}]{Casassus2012}
{Casassus}, S., {Perez M.}, S., {Jord{\'a}n}, A., {et~al.} 2012, \apjl, 754,
  L31

\bibitem[{{Christiaens} {et~al.}(2014){Christiaens}, {Casassus}, {Perez}, {van
  der Plas}, \& {M{\'e}nard}}]{Christiaens2014}
{Christiaens}, V., {Casassus}, S., {Perez}, S., {van der Plas}, G., \&
  {M{\'e}nard}, F. 2014, \apjl, 785, L12

\bibitem[{{Cohen}(1980)}]{Cohen1980}
{Cohen}, M. 1980, \aj, 85, 29

\bibitem[{{Connelley} {et~al.}(2007){Connelley}, {Reipurth}, \&
  {Tokunaga}}]{Connelley2007}
{Connelley}, M.~S., {Reipurth}, B., \& {Tokunaga}, A.~T. 2007, \aj, 133, 1528

\bibitem[{{Cuello} {et~al.}(2019){Cuello}, {Dipierro}, {Mentiplay}, {Price},
  {Pinte}, {Cuadra}, {Laibe}, {M{\'e}nard}, {Poblete}, \&
  {Montesinos}}]{Cuello2019}
{Cuello}, N., {Dipierro}, G., {Mentiplay}, D., {et~al.} 2019, \mnras, 483, 4114

\bibitem[{{Dong} {et~al.}(2015){Dong}, {Hall}, {Rice}, \& {Chiang}}]{Dong2015}
{Dong}, R., {Hall}, C., {Rice}, K., \& {Chiang}, E. 2015, \apjl, 812, L32

\bibitem[{{Dong} {et~al.}(2022){Dong}, {Liu}, {Cuello}, {Pinte},
  {{\'A}brah{\'a}m}, {Vorobyov}, {Hashimoto}, {K{\'o}sp{\'a}l}, {Chiang},
  {Takami}, {Chen}, {Dunham}, {Fukagawa}, {Green}, {Hasegawa}, {Henning},
  {Pavlyuchenkov}, {Pyo}, \& {Tamura}}]{Dong2022}
{Dong}, R., {Liu}, H.~B., {Cuello}, N., {et~al.} 2022, Nature Astronomy, 6, 331

\bibitem[{{Dullemond} {et~al.}(2019){Dullemond}, {K{\"u}ffmeier}, {Goicovic},
  {Fukagawa}, {Oehl}, \& {Kramer}}]{Dullemond2019}
{Dullemond}, C.~P., {K{\"u}ffmeier}, M., {Goicovic}, F., {et~al.} 2019, \aap,
  628, A20

\bibitem[{{Dunham} {et~al.}(2015){Dunham}, {Allen}, {Evans},
  {Broekhoven-Fiene}, {Cieza}, {Di Francesco}, {Gutermuth}, {Harvey},
  {Hatchell}, {Heiderman}, {Huard}, {Johnstone}, {Kirk}, {Matthews}, {Miller},
  {Peterson}, \& {Young}}]{Dunham2015}
{Dunham}, M.~M., {Allen}, L.~E., {Evans}, Neal~J., I., {et~al.} 2015, \apjs,
  220, 11

\bibitem[{{Garg} {et~al.}(2021){Garg}, {Pinte}, {Christiaens}, {Price},
  {Lazendic}, {Boehler}, {Casassus}, {Marino}, {Perez}, \& {Zuleta}}]{Garg2021}
{Garg}, H., {Pinte}, C., {Christiaens}, V., {et~al.} 2021, \mnras, 504, 782

\bibitem[{{Garufi} {et~al.}(2022){Garufi}, {Podio}, {Codella}, {Segura-Cox},
  {Vander Donckt}, {Mercimek}, {Bacciotti}, {Fedele}, {Kasper}, {Pineda},
  {Humphreys}, \& {Testi}}]{Garufi2022}
{Garufi}, A., {Podio}, L., {Codella}, C., {et~al.} 2022, \aap, 658, A104

\bibitem[{{Ginski} {et~al.}(2021){Ginski}, {Facchini}, {Huang}, {Benisty},
  {Vaendel}, {Stapper}, {Dominik}, {Bae}, {M{\'e}nard}, {Muro-Arena},
  {Hogerheijde}, {McClure}, {van Holstein}, {Birnstiel}, {Boehler}, {Bohn},
  {Flock}, {Mamajek}, {Manara}, {Pinilla}, {Pinte}, \& {Ribas}}]{Ginski2021}
{Ginski}, C., {Facchini}, S., {Huang}, J., {et~al.} 2021, \apjl, 908, L25

\bibitem[{{Hall} {et~al.}(2020){Hall}, {Dong}, {Teague}, {Terry}, {Pinte},
  {Paneque-Carre{\~n}o}, {Veronesi}, {Alexander}, \& {Lodato}}]{Hall2020}
{Hall}, C., {Dong}, R., {Teague}, R., {et~al.} 2020, \apj, 904, 148

\bibitem[{{Harsono} {et~al.}(2011){Harsono}, {Alexander}, \&
  {Levin}}]{Harsono2011}
{Harsono}, D., {Alexander}, R.~D., \& {Levin}, Y. 2011, \mnras, 413, 423

\bibitem[{{Hennebelle} {et~al.}(2017){Hennebelle}, {Lesur}, \&
  {Fromang}}]{Hennebelle2017}
{Hennebelle}, P., {Lesur}, G., \& {Fromang}, S. 2017, \aap, 599, A86

\bibitem[{{Huang} {et~al.}(2020){Huang}, {Andrews}, {{\"O}berg}, {Ansdell},
  {Benisty}, {Carpenter}, {Isella}, {P{\'e}rez}, {Ricci}, {Williams}, {Wilner},
  \& {Zhu}}]{Huang2020}
{Huang}, J., {Andrews}, S.~M., {{\"O}berg}, K.~I., {et~al.} 2020, \apj, 898,
  140

\bibitem[{{Huang} {et~al.}(2021){Huang}, {Bergin}, {{\"O}berg}, {Andrews},
  {Teague}, {Law}, {Kalas}, {Aikawa}, {Bae}, {Bergner}, {Booth}, {Bosman},
  {Calahan}, {Cataldi}, {Cleeves}, {Czekala}, {Ilee}, {Le Gal}, {Guzm{\'a}n},
  {Long}, {Loomis}, {M{\'e}nard}, {Nomura}, {Qi}, {Schwarz}, {Tsukagoshi},
  {van't Hoff}, {Walsh}, {Wilner}, {Yamato}, \& {Zhang}}]{Huang2021}
{Huang}, J., {Bergin}, E.~A., {{\"O}berg}, K.~I., {et~al.} 2021, \apjs, 257, 19

\bibitem[{{Huang} {et~al.}(2022){Huang}, {Ginski}, {Benisty}, {Ren}, {Bohn},
  {Choquet}, {{\"O}berg}, {Ribas}, {Bae}, {Bergin}, {Birnstiel}, {Boehler},
  {Facchini}, {Harsono}, {Hogerheijde}, {Long}, {Manara}, {M{\'e}nard},
  {Pinilla}, {Pinte}, {Rab}, {Williams}, \& {Zurlo}}]{Huang2022}
{Huang}, J., {Ginski}, C., {Benisty}, M., {et~al.} 2022, \apj, 930, 171

\bibitem[{{Hubble}(1922)}]{Hubble1922}
{Hubble}, E.~P. 1922, \apj, 56, 400

\bibitem[{{Hunziker} {et~al.}(2021){Hunziker}, {Schmid}, {Ma}, {Menard},
  {Avenhaus}, {Boccaletti}, {Beuzit}, {Chauvin}, {Dohlen}, {Dominik}, {Engler},
  {Ginski}, {Gratton}, {Henning}, {Langlois}, {Milli}, {Mouillet}, {Tschudi},
  {van Holstein}, \& {Vigan}}]{Hunziker2021}
{Hunziker}, S., {Schmid}, H.~M., {Ma}, J., {et~al.} 2021, \aap, 648, A110

\bibitem[{{Joy}(1945)}]{Joy1945}
{Joy}, A.~H. 1945, \apj, 102, 168

\bibitem[{{Kuffmeier} {et~al.}(2021){Kuffmeier}, {Dullemond}, {Reissl}, \&
  {Goicovic}}]{Kuffmeier2021}
{Kuffmeier}, M., {Dullemond}, C.~P., {Reissl}, S., \& {Goicovic}, F.~G. 2021,
  \aap, 656, A161

\bibitem[{{Kuffmeier} {et~al.}(2018){Kuffmeier}, {Frimann}, {Jensen}, \&
  {Haugb{\o}lle}}]{Kuffmeier2018}
{Kuffmeier}, M., {Frimann}, S., {Jensen}, S.~S., \& {Haugb{\o}lle}, T. 2018,
  \mnras, 475, 2642

\bibitem[{{Kuffmeier} {et~al.}(2020){Kuffmeier}, {Goicovic}, \&
  {Dullemond}}]{Kuffmeier2020}
{Kuffmeier}, M., {Goicovic}, F.~G., \& {Dullemond}, C.~P. 2020, \aap, 633, A3

\bibitem[{{Kuffmeier} {et~al.}(2017){Kuffmeier}, {Haugb{\o}lle}, \&
  {Nordlund}}]{Kuffmeier2017}
{Kuffmeier}, M., {Haugb{\o}lle}, T., \& {Nordlund}, {\r{A}}. 2017, \apj, 846, 7

\bibitem[{{Kurtovic} {et~al.}(2018){Kurtovic}, {P{\'e}rez}, {Benisty}, {Zhu},
  {Zhang}, {Huang}, {Andrews}, {Dullemond}, {Isella}, {Bai}, {Carpenter},
  {Guzm{\'a}n}, {Ricci}, \& {Wilner}}]{Kurtovic2018}
{Kurtovic}, N.~T., {P{\'e}rez}, L.~M., {Benisty}, M., {et~al.} 2018, \apjl,
  869, L44

\bibitem[{{Kuznetsova} {et~al.}(2022){Kuznetsova}, {Bae}, {Hartmann}, \& {Mac
  Low}}]{Kuznetsova2022}
{Kuznetsova}, A., {Bae}, J., {Hartmann}, L., \& {Mac Low}, M.-M. 2022, \apj,
  928, 92

\bibitem[{{Kuznetsova} {et~al.}(2020){Kuznetsova}, {Hartmann}, \&
  {Heitsch}}]{Kuznetsova2020}
{Kuznetsova}, A., {Hartmann}, L., \& {Heitsch}, F. 2020, \apj, 893, 73

\bibitem[{{Magakian}(2003)}]{Magakian2003}
{Magakian}, T.~Y. 2003, \aap, 399, 141

\bibitem[{{Manara} {et~al.}(2018){Manara}, {Morbidelli}, \&
  {Guillot}}]{Manara2018}
{Manara}, C.~F., {Morbidelli}, A., \& {Guillot}, T. 2018, \aap, 618, L3

\bibitem[{{Marton} {et~al.}(2016){Marton}, {T{\'o}th}, {Paladini}, {Kun},
  {Zahorecz}, {McGehee}, \& {Kiss}}]{Marton2016}
{Marton}, G., {T{\'o}th}, L.~V., {Paladini}, R., {et~al.} 2016, \mnras, 458,
  3479

\bibitem[{{McKee} \& {Ostriker}(2007)}]{McKee2007}
{McKee}, C.~F. \& {Ostriker}, E.~C. 2007, \araa, 45, 565

\bibitem[{{Mulders} {et~al.}(2021){Mulders}, {Pascucci}, {Ciesla}, \&
  {Fernandes}}]{Mulders2021}
{Mulders}, G.~D., {Pascucci}, I., {Ciesla}, F.~J., \& {Fernandes}, R.~B. 2021,
  \apj, 920, 66

\bibitem[{{Murillo} {et~al.}(2022){Murillo}, {van Dishoeck}, {Hacar},
  {Harsono}, \& {J{\o}rgensen}}]{Murillo2022}
{Murillo}, N.~M., {van Dishoeck}, E.~F., {Hacar}, A., {Harsono}, D., \&
  {J{\o}rgensen}, J.~K. 2022, \aap, 658, A53

\bibitem[{{Nanne} {et~al.}(2019){Nanne}, {Nimmo}, {Cuzzi}, \&
  {Kleine}}]{Nanne2019N}
{Nanne}, J. A.~M., {Nimmo}, F., {Cuzzi}, J.~N., \& {Kleine}, T. 2019, Earth and
  Planetary Science Letters, 511, 44

\bibitem[{{Padoan} {et~al.}(2014){Padoan}, {Haugb{\o}lle}, \&
  {Nordlund}}]{Padoan2014}
{Padoan}, P., {Haugb{\o}lle}, T., \& {Nordlund}, {\r{A}}. 2014, \apj, 797, 32

\bibitem[{{Padoan} {et~al.}(2005){Padoan}, {Kritsuk}, {Norman}, \&
  {Nordlund}}]{Padoan2005}
{Padoan}, P., {Kritsuk}, A., {Norman}, M.~L., \& {Nordlund}, {\r{A}}. 2005,
  \apjl, 622, L61

\bibitem[{{Pineda} {et~al.}(2020){Pineda}, {Segura-Cox}, {Caselli},
  {Cunningham}, {Zhao}, {Schmiedeke}, {Maureira}, \& {Neri}}]{Pineda2020}
{Pineda}, J.~E., {Segura-Cox}, D., {Caselli}, P., {et~al.} 2020, Nature
  Astronomy, 4, 1158

\bibitem[{{Rodriguez} {et~al.}(2018){Rodriguez}, {Loomis}, {Cabrit}, {Haworth},
  {Facchini}, {Dougados}, {Booth}, {Jensen}, {Clarke}, {Stassun}, {Dent}, \&
  {Pety}}]{Rodriguez2018}
{Rodriguez}, J.~E., {Loomis}, R., {Cabrit}, S., {et~al.} 2018, \apj, 859, 150

\bibitem[{{Schneider} {et~al.}(2003){Schneider}, {Wood}, {Silverstone},
  {Hines}, {Koerner}, {Whitney}, {Bjorkman}, \& {Lowrance}}]{Schneider2003}
{Schneider}, G., {Wood}, K., {Silverstone}, M.~D., {et~al.} 2003, \aj, 125,
  1467

\bibitem[{{Stolker} {et~al.}(2016){Stolker}, {Dominik}, {Min}, {Garufi},
  {Mulders}, \& {Avenhaus}}]{Stolker2016}
{Stolker}, T., {Dominik}, C., {Min}, M., {et~al.} 2016, \aap, 596, A70

\bibitem[{{Tang} {et~al.}(2012){Tang}, {Guilloteau}, {Pi{\'e}tu}, {Dutrey},
  {Ohashi}, \& {Ho}}]{Tang2012}
{Tang}, Y.~W., {Guilloteau}, S., {Pi{\'e}tu}, V., {et~al.} 2012, \aap, 547, A84

\bibitem[{{Thieme} {et~al.}(2022){Thieme}, {Lai}, {Lin}, {Cheong}, {Lee},
  {Yen}, {Li}, {Lam}, \& {Zhao}}]{Thieme2022}
{Thieme}, T.~J., {Lai}, S.-P., {Lin}, S.-J., {et~al.} 2022, \apj, 925, 32

\bibitem[{{Thies} {et~al.}(2011){Thies}, {Kroupa}, {Goodwin}, {Stamatellos}, \&
  {Whitworth}}]{Thies2011}
{Thies}, I., {Kroupa}, P., {Goodwin}, S.~P., {Stamatellos}, D., \& {Whitworth},
  A.~P. 2011, \mnras, 417, 1817

\bibitem[{{Throop} \& {Bally}(2008)}]{Throop2008}
{Throop}, H.~B. \& {Bally}, J. 2008, \aj, 135, 2380

\bibitem[{{Toomre}(1964)}]{Toomre1964}
{Toomre}, A. 1964, \apj, 139, 1217

\bibitem[{{Valdivia-Mena} {et~al.}(2022){Valdivia-Mena}, {Pineda},
  {Segura-Cox}, {Caselli}, {Neri}, {L{\'o}pez-Sepulcre}, {Cunningham},
  {Bouscasse}, {Semenov}, {Henning}, {Pi{\'e}tu}, {Chapillon}, {Dutrey},
  {Fuente}, {Guilloteau}, {Hsieh}, {Jim{\'e}nez-Serra}, {Marino}, {Maureira},
  {Smirnov-Pinchukov}, {Tafalla}, \& {Zhao}}]{Valdivia-Men2022}
{Valdivia-Mena}, M.~T., {Pineda}, J.~E., {Segura-Cox}, D.~M., {et~al.} 2022,
  arXiv e-prints, arXiv:2208.01023

\bibitem[{{Vorobyov} {et~al.}(2020){Vorobyov}, {Skliarevskii}, {Elbakyan},
  {Takami}, {Liu}, {Liu}, \& {Akiyama}}]{Vorobyov2020}
{Vorobyov}, E.~I., {Skliarevskii}, A.~M., {Elbakyan}, V.~G., {et~al.} 2020,
  \aap, 635, A196

\bibitem[{{Yen} {et~al.}(2019){Yen}, {Gu}, {Hirano}, {Koch}, {Lee}, {Liu}, \&
  {Takakuwa}}]{Yen2019}
{Yen}, H.-W., {Gu}, P.-G., {Hirano}, N., {et~al.} 2019, \apj, 880, 69

\bibitem[{{Zapata} {et~al.}(2020){Zapata}, {Rodr{\'\i}guez},
  {Fern{\'a}ndez-L{\'o}pez}, {Palau}, {Estalella}, {Osorio}, {Anglada}, \&
  {Huelamo}}]{Zapata2020}
{Zapata}, L.~A., {Rodr{\'\i}guez}, L.~F., {Fern{\'a}ndez-L{\'o}pez}, M.,
  {et~al.} 2020, \apj, 896, 132

\bibitem[{{Zucker} {et~al.}(2020){Zucker}, {Speagle}, {Schlafly}, {Green},
  {Finkbeiner}, {Goodman}, \& {Alves}}]{Zucker2020}
{Zucker}, C., {Speagle}, J.~S., {Schlafly}, E.~F., {et~al.} 2020, \aap, 633,
  A51

\end{thebibliography}

\newpage

\begin{appendix}

\section{Star-forming regions} \label{app:sfrs}
\begin{figure*}[h!]
\centering
\includegraphics[width=0.7\linewidth]{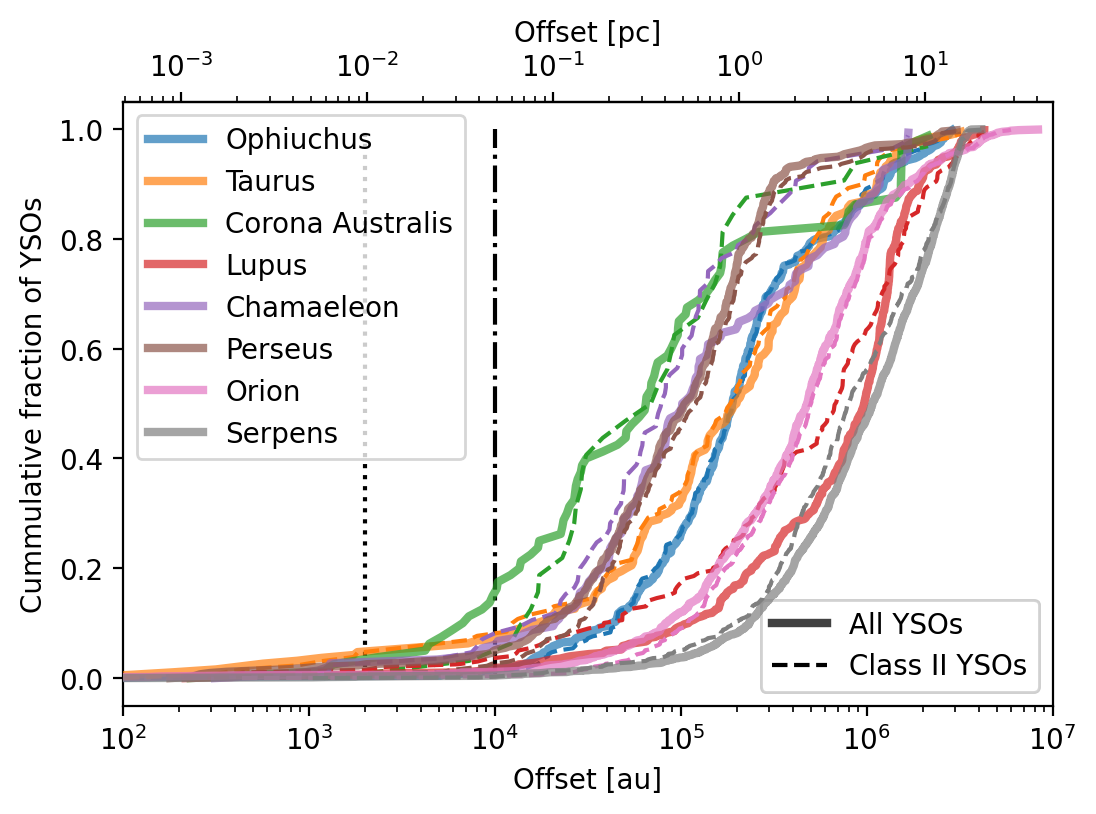}
\caption{Cumulative distribution of the fraction of YSOs with distance to the nearest RNe less than the given offset, as discussed in Section \ref{sec:discussion_RNe}, for different SFRs. Solid lines represent all the YSOs and dashed lines represent only Class II sources. Vertical dotted and dash-dotted lines denote offset values of 2000 and 10000 au, respectively.}
\label{fig:OffsetDist}
\end{figure*}

Figure \ref{fig:DSSmaps} shows DSS optical images of all the SFRs listed in Table \ref{tab:sfrs}, as discussed in Section \ref{sec:RNesources}.
Figure \ref{fig:OffsetDist} shows fraction of all YSOs (solid lines) and just Class II sources (dashed lines) which are associated with RNe as a function of offset thresholds used to define association, as discussed in Section \ref{sec:discussion_RNe}.

\FloatBarrier

\begin{strip}
    \centering
    \includegraphics[width=1\linewidth]{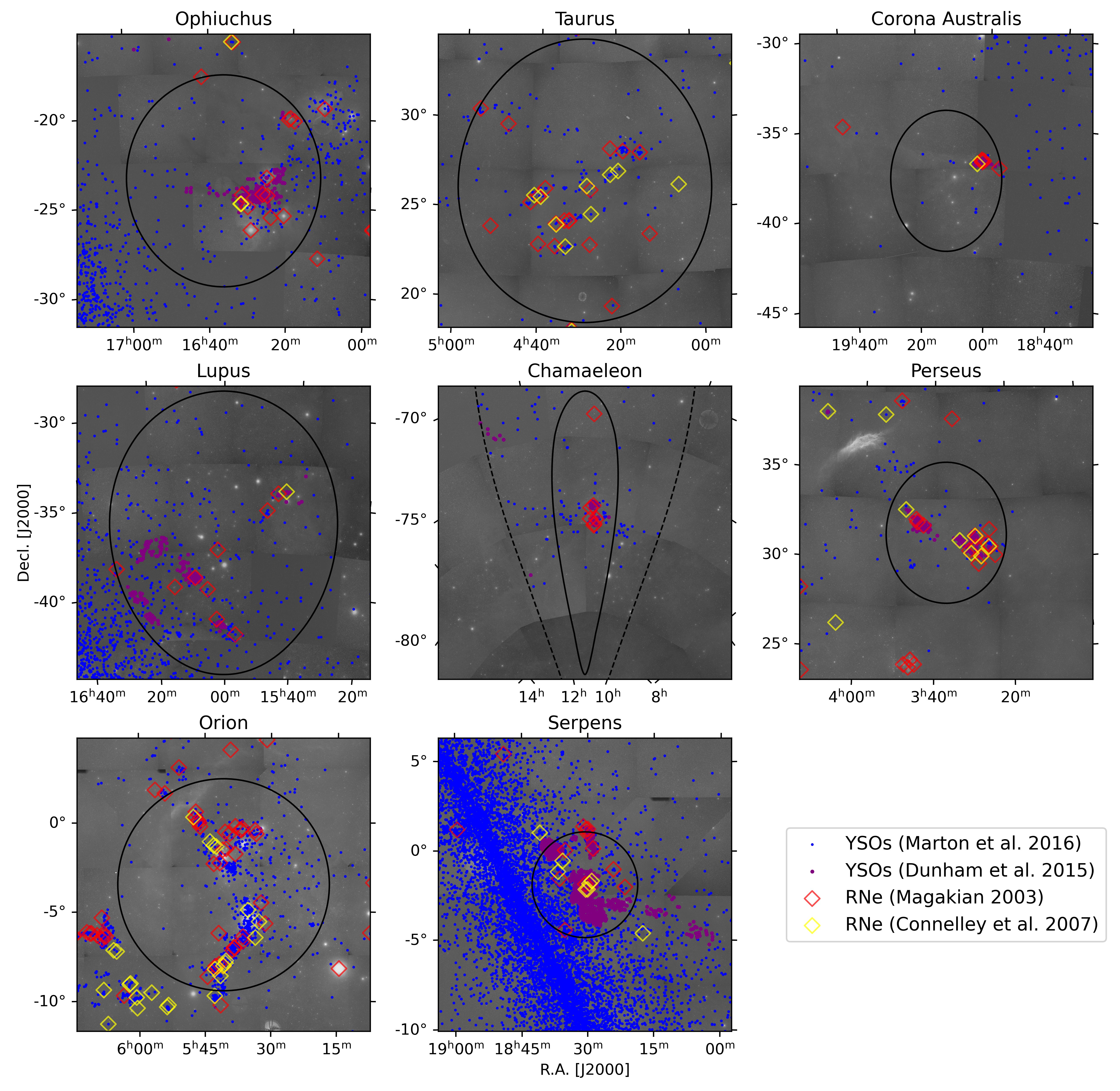}
    \captionof{figure}{DSS optical images of all the SFRs listed in Table \ref{tab:sfrs}. Solid-black curves denotes circular boundaries of these SFRs, as parameterized by the "Radius" column of Table \ref{tab:sfrs}.
    Blue and purple circles represent YSOs from \citet{Marton2016} and \citet{Dunham2015} catalogues, respectively.
    Red and yellow open diamonds represent RNe from \citet{Magakian2003} and \citet{Connelley2007} catalogues, respectively.
    Dashed-black curve in Chamaeleon's map denote a circle with radius of $20^{\circ}$.}
    \label{fig:DSSmaps}
\end{strip}


\hspace{7pt}



\FloatBarrier

\section{Distribution of spectral indices} \label{app:alpha_dist}
Figure \ref{fig:alphas} shows the distribution of extinction-corrected spectral indices ($\alpha$', solid bars) and originally measured spectral indices ($\alpha$, dashed-grey line), as discussed in Section \ref{sec:RNesources}. The distribution of $\alpha$ values are shifted to the right because foreground extinction can artificially increase the observed infrared excess for a source.



\begin{strip}
    \centering
    \includegraphics[width=0.7\linewidth]{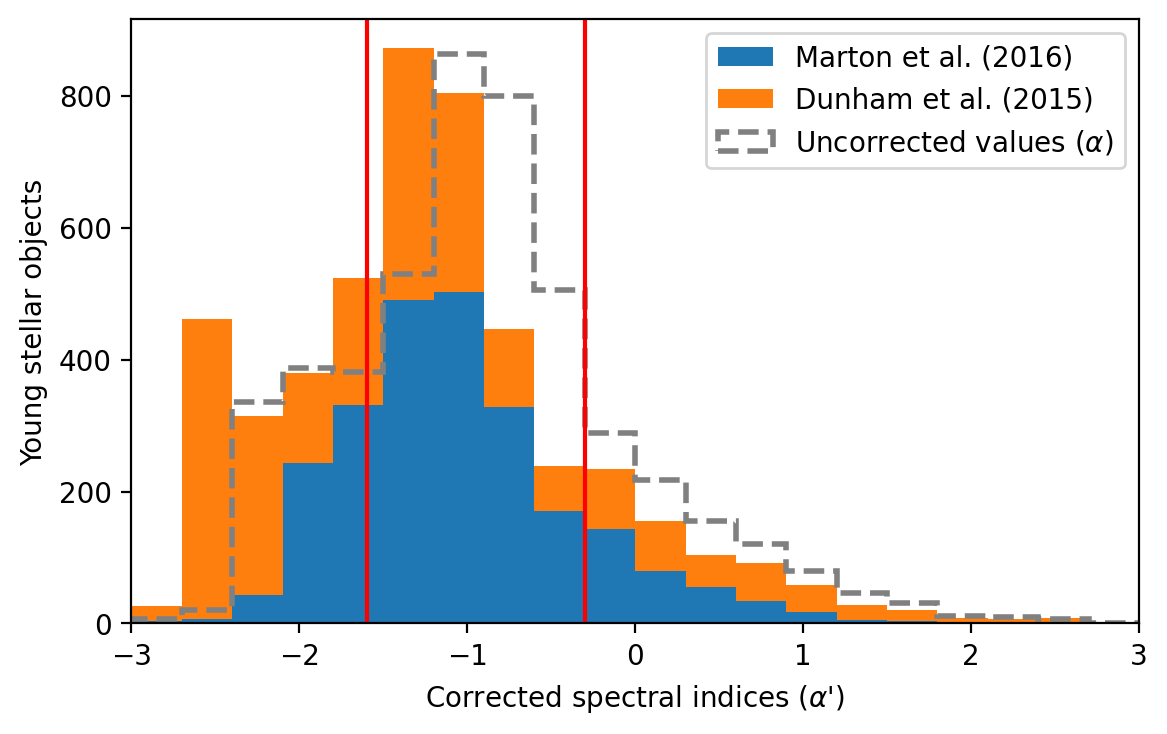}
    \label{fig:alphas}
    \captionof{figure}{Distribution of extinction-corrected infrared spectral indices ($\alpha$', solid bars) and measured spectral indices ($\alpha$, grey-dashed steps) for all the 4930 YSOs in SFRs, as discussed in Section \ref{sec:discussion_RNe}. Blue bars denote $\alpha$' values estimated for sources exclusively from \citet{Marton2016}. Orange bars denote $\alpha$' values for sources from \citet{Dunham2015}. Red vertical lines mark the range of values for a YSO to be classified as a Class II source ($-1.6 \leq \alpha' < -0.3$). }
\end{strip}

\FloatBarrier
\newpage
\hspace{4pt}

\section{Class II sources near RNe}  \label{app:table}
Table \ref{tab:class2} gives coordinates (first two columns), SIMBAD identifiers (third column), the SFR (fourth column), spectral indices (fifth and sixth columns), and RNe catalogue identifiers (last two columns) for all the Class II sources in the vicinity of RNe, as discussed in Section \ref{sec:RNesources}.

\begin{table}[H]
\onecolumn
\caption{Class II YSOs associated with RNe}
\begin{tabular}{cccccccc} 
\hline\hline
R.A. [$^{\circ}$]& Decl. [$^{\circ}$]& Simbad Id. & Region & $\alpha$ & $\alpha$' & Magakian RNe Id. & Connelley RNe Id. \\
\hline
247.96698 & -24.93782 & ISO-Oph 204 & Ophiuchus & -0.17 & -0.51 & - & 66 \\
239.17449 & -42.32318 & HD 142527 & Lupus & -0.6 & -0.91 & 641 & - \\
236.30347 & -34.29186 & CD-33 10685 & Lupus & -0.64 & -0.95 & 634 & - \\
237.02178 & -35.26469 & V* HN Lup & Lupus & -0.84 & -1.15 & 636 & - \\
277.19941 & 0.14439 & V* VV Ser & Serpens & -0.78 & -1.05 & 766 & - \\
85.20028 & -8.09964 & CoKu DL Ori G1 & Orion & -0.93 & -1.24 & 132 & - \\
167.01364 & -77.65476 & HD  97048b & Chamaeleon & -0.07 & -0.38 & 533 & - \\
168.11282 & -76.73947 & BRAN 341D & Chamaeleon & -0.76 & -1.11 & 545 & - \\
168.12797 & -76.73998 & V* CW Cha & Chamaeleon & -0.61 & -1.09 & 545 & - \\
285.28588 & -36.95575 & V* S CrA B & Corona Australis & -0.8 & -1.22 & 781 & - \\
68.13232 & 24.33411 & V* FZ Tau & Taurus & -0.86 & -1.17 & 74 & - \\
68.39192 & 24.35472 & V* GI Tau & Taurus & -0.62 & -0.93 & 75 & - \\
68.12742 & 24.33257 & V* FY Tau & Taurus & -1.12 & -1.43 & 74 & - \\
68.39412 & 24.35176 & V* GK Tau & Taurus & -0.67 & -0.98 & 75 & - \\
69.61912 & 26.1804 & V* DO Tau & Taurus & -0.51 & -0.82 & 78 & - \\
68.92066 & 24.18566 & NAME CoKu Tau 3 & Taurus & -0.9 & -1.21 & 76 & - \\
68.97004 & 22.90634 & V* HP Tau & Taurus & -0.57 & -0.88 & 77 & - \\
55.73321 & 31.97828 & 2MASS J03425596+3158419 & Perseus & -0.98 & -0.84 & 48 & - \\
52.68335 & 30.54634 & EM* LkHA  326 & Perseus & -0.64 & -0.89 & 45 & - \\
52.21743 & 30.75151 & EM* LkHA  325 & Perseus & -0.69 & -0.78 & 41 & - \\
235.755 & -34.15417 & HH 185 & Lupus & -0.21 & -0.31 & - & 64 \\
\hline
\label{tab:class2}
\end{tabular}
\tablefoot{List of 21 Class II YSOs ($-1.6 \leq \alpha' < -0.3$) associated with RNe (distance to nearest RNe $\lesssim$ 2000~au), as discussed in Section \ref{sec:RNesources}. $\alpha$ and $\alpha$' values are measured and extinction-corrected spectral indices, respectively. Last two columns show index numbers for matched RNe in \citet{Magakian2003} and \citet{Connelley2007} catalogues.}
\end{table}
\twocolumn

\section{Required observations and analysis}
\label{sec:discussion_future}

In order to further test a possible link between RNe and late infall, a deep uniform survey of large-scale structures is needed for Class II sources associated with RNe, as suggested in Section \ref{sec:discussion_RNe}. Ideal observational parameters for such a survey are discussed below.

For what concerns the angular scales, both observations (Figure~\ref{fig:known_sources}) and simulations \citep[e.g.][]{Kuffmeier2020} suggest that the infalling streamers should be roughly kilo-au scales in length. Therefore, observations needed to study these structures should have a large enough maximum recoverable angular scale ($\gtrsim1000$~au), so as to not filter out large-scale emission. For the typical distance of 150 pc to nearby SFRs, this physical scale corresponds to the largest angular scale of $\gtrsim7\arcsec$. On the other hand, spatial resolution of such observations should be roughly $\lesssim100$~au ($\lesssim0.7\arcsec$ at a distance of 150~pc), in order to resolve the connection between large-scale structures and protoplanetary disks. Such a resolution should also be adequate to resolve the width of infalling streamers (Figure \ref{fig:known_sources}).

 In terms of spectral resolution, free-fall velocity for the infall of material can be estimated as $v = \sqrt{2GM_{*}/R}$, where $G$ is the gravitational constant, $M_{*}$ is the stellar mass, and $R$ is the free-fall length scale. For the typical stellar mass of $\sim0.5 M_{\odot}$ and expected infall length scale of $\sim1,000$~au, the free-fall velocity should be $\sim0.95$~km~s$^{-1}$. Assuming we see such an infalling streamer at an intermediate inclination of $45^{\circ}$, observed velocity difference would be $\sim0.65$~km~s$^{-1}$. In order to resolve the velocity profile, we would need at least three independent data points, and thus a spectral resolution of $\lesssim0.2$~km~s$^{-1}$. 

The sensitivity requirements of the ideal observations can be based on the past observations of such large-scale structures. Among the five sources discussed in Section \ref{sec:knownsources}, AB Aur and SU Aur are exceptionally bright and may not be representatives for the overall sample. For RU Lup, the signal-to-noise ratio for the spiral structures was sub-optimal ($\lesssim3$) in the individual channel \citep[see Figure 5,][]{Huang2020}, which can make it hard to study the background dynamical processes. Thus, sensitivity requirements of the observations can be based on observations of GM Aur and DO Tau, and for both of which the brightness-temperature sensitivity was $\sim250$~mK (normalised to a channel width of $0.2$~km~s$^{-1}$).

If large-scale structures are observed around other Class II sources, gas kinematics can be analysed to understand the dominant dynamical processes. A first step could be to check if the material is gravitationally bound to the protostellar system. For this the kinetic energy can be computed along the streamer, using the relative line-of-sight velocities, and compared to gravitational energy, similar to the analysis done for DO Tau by \citet{Huang2022} (see Figure 12). Furthermore, position-velocity diagrams, along any detected streamer, can be modelled and compared to the velocity profiles expected for different kinematic features such as rotation ($v \propto R^{-1}$, for conserved angular momentum) and infall ($v \propto R^{-0.5}$, for free fall), similar to the analysis done for less evolved protostars HL Tau \citep{Yen2019} and Lupus 3-MMS \citep{Thieme2022}. 

Another way to infer late infall could be to study gas kinematics together with NIR polarisation observations, as was done for SU Aur by \citet{Ginski2021}. The degree of polarisation in such observations can be correlated to the dust scattering angles, which are expected to depend on the three-dimensional morphology of dust structures \citep[e.g.][]{Stolker2016}. Studying the morphology and gas kinematics in larger-scale ($\sim10,000$~au) clouds can also allow us to judge the possibility of late infall \citep{Tang2012, Dullemond2019}.

Finally, late infall can also be inferred by observing these systems using different chemical species. Though CO has a high surface brightness, making it ideal to detect faint structures, it is also likely to be polluted by the emission from diffuse gas in these clouds. For less evolved sources, infalling streamers have also been observed in tracers such as HCO$^{+}$, HC$_3$N, HC$_5$N, CCS, $^{13}$CS, HNC, and H$_2$CO \citep{Yen2019,Pineda2020,Murillo2022, Valdivia-Men2022}. Moreover, material falling onto protoplanetary disks also creates shocks, which can be observed using shock tracers such as SiO, SO, and SO$_2$ \citep[e.g.][]{Garufi2022}. A dedicated chemical study of streamers could also allow us to identify better chemical tracers for these structures for a large-scale survey.

\section{Alternative explanations for large-scale structures}
\label{sec:discussion_other}

The large-scale CO structures discussed in Section \ref{sec:knownsources} and \ref{sec:RNesources} could also be due to other dynamical processes besides late infall \citep[e.g.][]{Huang2020}. One of the other prominent causes, particularly for spiral-like structures, could be a tidal interaction of stellar companions, as it has been observed in some other multiple systems \citep[e.g.][]{Rodriguez2018,Kurtovic2018,Zapata2020}. 
The two sources we found with large-scale structures, HD 142527 and S CrA (Section \ref{sec:RNesources}), are binaries, and thus some of the structures we observe around them (Figure \ref{fig:rne_sources}, panel a and c) could be due to tidal interactions between protostars and surrounding gas. 

Furthermore, such structures can also be created due to close encounters by neighbouring YSOs, as predicted by several hydrodynamic simulations \citep[e.g.][]{Cuello2019,Vorobyov2020} and likely observed for a few sources \citep[e.g.][]{Dong2022}. The role of these stellar flybys can be checked by looking at relative distances and velocities of nearby YSOs, as was done for SU Aur \citep[Appendix D]{Ginski2021}. 

Gravitational instabilities can be another possible way to form spiral-like structures, if the disks are massive enough \citep[e.g.][]{Dong2015}, as generally inferred by Toomre's Q parameter \citep{Toomre1964}. Such instabilities are expected to leave characteristic 'wiggle' signatures in the gas kinematics, which can be used to identify them \citep{Hall2020}. Moreover, \citet{Harsono2011} showed that such instabilities can also be triggered by the infall of material. Irrespective of the cause of such non-Keplerian structures, both approaches followed in Sec. \ref{sec:knownsources} and in Sec. \ref{sec:RNesources} suggest that the vicinity of a RN can be an effective criterion to identify Class II disks that present large-scale structures.

\section{Channel maps} \label{app:channel_maps}
\begin{figure*}[h]
\centering
\includegraphics[width=0.85\linewidth]{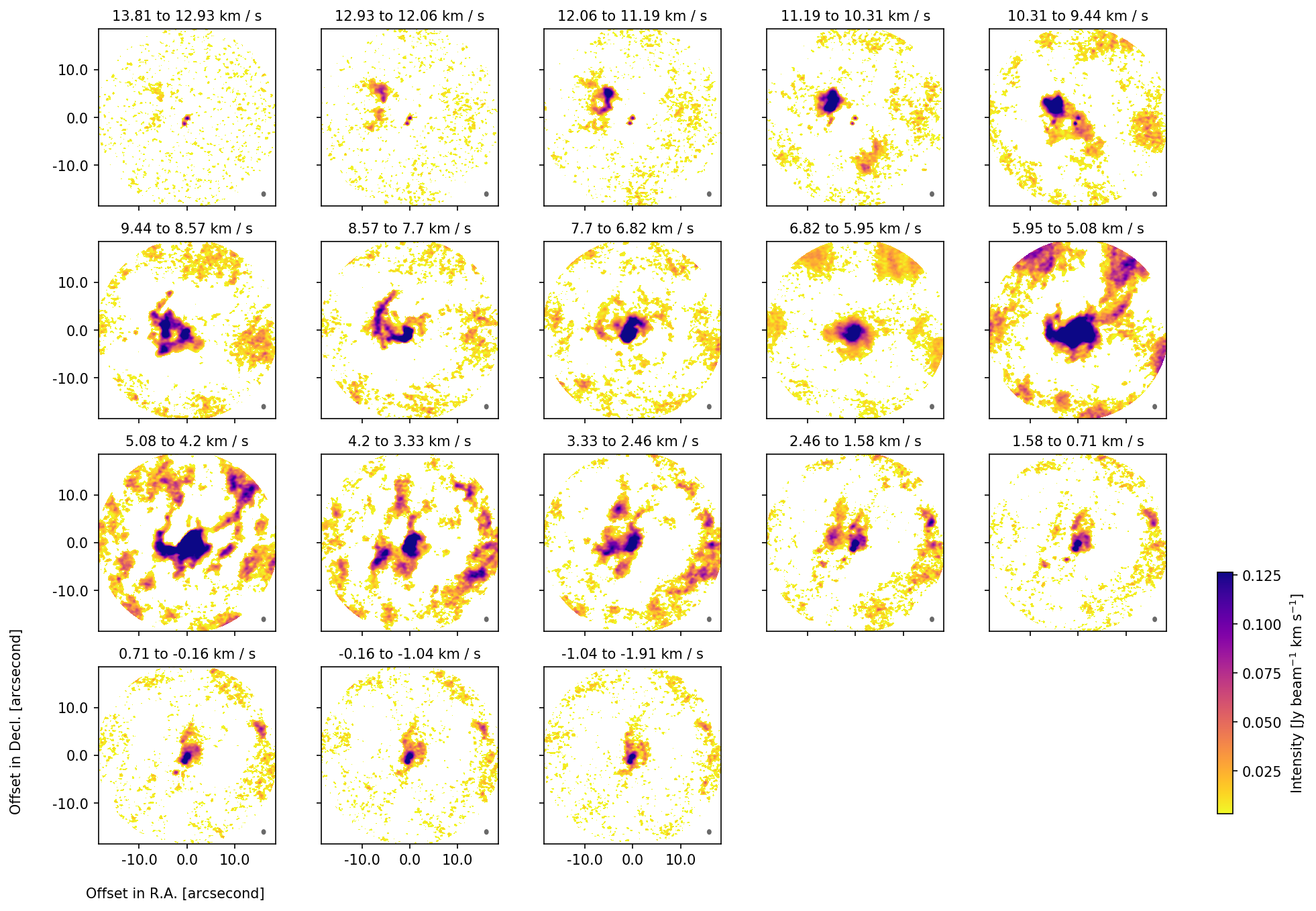}
\caption{ALMA $^{12}$CO (2--1) channel maps for S CrA archival observations (Project code: 2019.1.01792.S). Emission only from pixels with an intensity $>2\sigma$ was considered. Grey ellipses in the bottom right corners of the maps represent the beam size.}
\label{fig:scra_channels}
\end{figure*}

\begin{figure*}[h]
\centering
\includegraphics[width=0.85\linewidth]{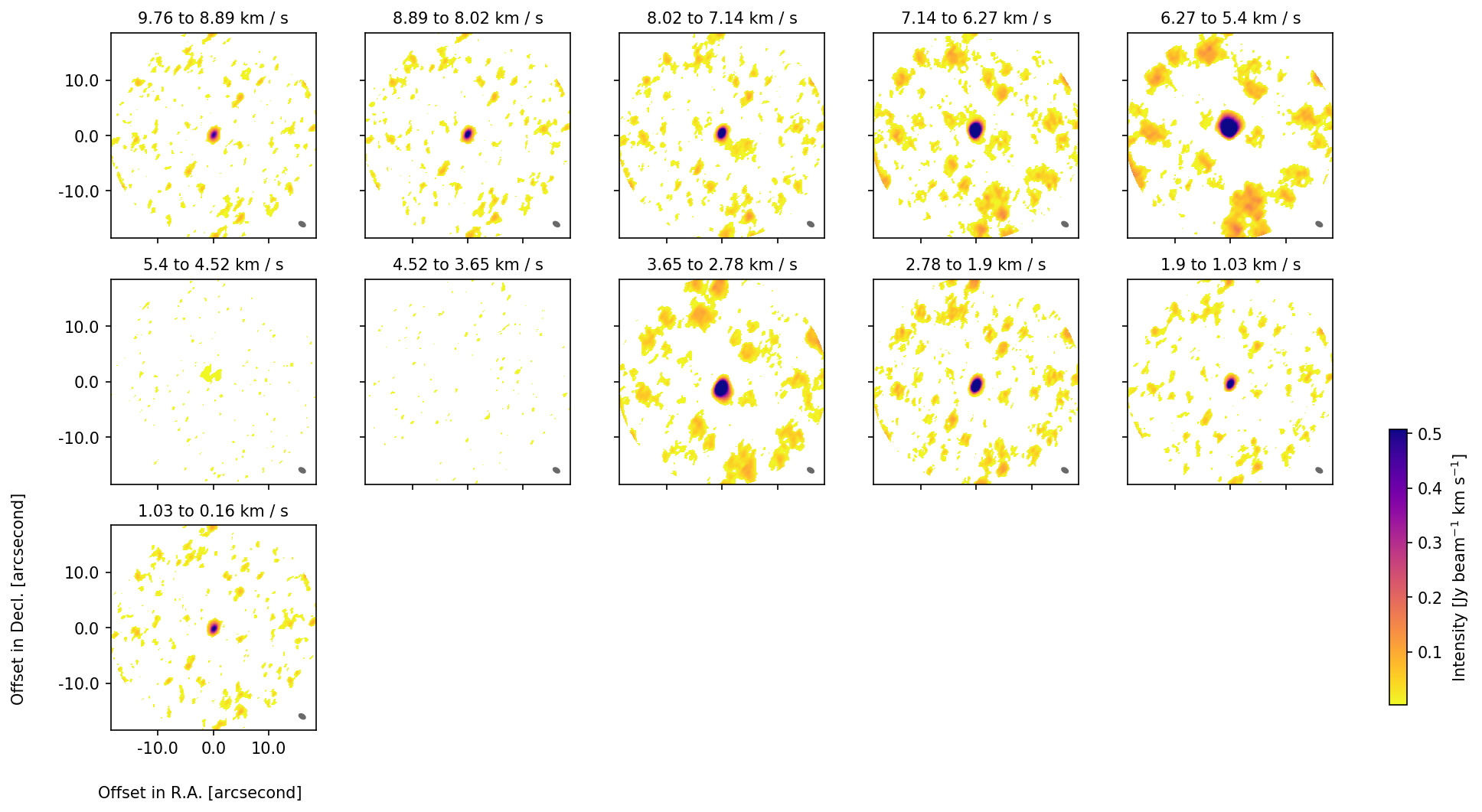}
\caption{ALMA $^{12}$CO (2--1) channel maps for HD 97048 archival observations (Project code: 2015.1.00192.S). Emission only from pixels with an intensity $>2\sigma$ was considered. Grey ellipses in bottom right corners of the
maps represent the beam size.}
\label{fig:hd97048_channels}
\end{figure*}

Figure \ref{fig:scra_channels} and \ref{fig:hd97048_channels} show channels maps of S CrA and HD 97048, respectively, as discussed in Section \ref{sec:RNesources}.

\end{appendix}

\end{document}